\documentstyle[prl,aps,epsfig,floats]{revtex}

\def\be{\begin{equation}}
\def\ee{\end{equation}}
\def\ba{\begin{eqnarray}}
\def\ea{\end{eqnarray}}

\begin{document}

\twocolumn[\hsize\textwidth\columnwidth\hsize\csname
@twocolumnfalse\endcsname

\preprint{hep-th/0110260}
\draft
\tighten

\title{A rotating black ring in five dimensions}
\author{Roberto Emparan$^a$\cite{aff}
and Harvey S. Reall$^b$}
\address{$^a$ Theory Division, CERN,
CH-1211 Geneva 23, Switzerland\\
$^b$ Physics Department, Queen Mary College,
Mile End Road, London E1 4NS, United Kingdom}
\setlength{\footnotesep}{0.5\footnotesep}
\maketitle

\begin{abstract}
The vacuum Einstein equations in five dimensions are shown to admit a
solution describing a stationary asymptotically flat spacetime regular on and
outside an event horizon of topology $S^1 \times S^2$. It describes a
rotating ``black ring''. This is the first example of a stationary
asymptotically flat vacuum solution with an event horizon 
of non-spherical topology.
There is a range of values for the mass and angular momentum for which
there exist two black ring solutions as well as a black hole solution.
Therefore the uniqueness theorems valid in four dimensions do not have
simple five-dimensional generalizations. It is suggested that
increasing the spin of a five-dimensional black hole beyond a critical
value results in a transition to a black ring, which can have an
arbitrarily large angular momentum for a given mass.
\end{abstract}

\vskip2pc]



Black holes in four spacetime dimensions are highly constrained
objects. A number of classical theorems show that a stationary, 
asymptotically flat, vacuum black hole is completely characterized by
its mass and spin \cite{unique}, and event horizons of non-spherical
topology are forbidden \cite{topcensor}.

In this paper we show explicitly that in five dimensions the situation
cannot be so simple by exhibiting an asymptotically flat, stationary,
vacuum solution with a horizon of topology $S^1 \times S^2$: a black
ring. The ring rotates along the $S^1$ and this balances its
gravitational self-attraction. The solution is characterized by its
mass $M$ and spin $J$. The black hole of \cite{myers:86} with rotation
in a single plane (and horizon of topology $S^3$) can be obtained as a
branch of the same family of solutions. We will show that there exist
black holes and black rings with the same values of $M$ and $J$. They
can be distinguished by their topology, and by their mass-dipole
measured at infinity. This shows that there is no obvious
five-dimensional analogue of the uniqueness theorems.

$S^1 \times S^2$ is one of the few possible topologies for the event
horizon in five dimensions that was not ruled out by the analysis in
\cite{cai:01} (although this argument does not apply directly to our
black ring because it assumes time symmetry). An explicit solution with
a regular (but degenerate) horizon of topology $S^1\times S^2$, and
spacelike infinity with $S^3$ topology has been built recently in
\cite{emparan:01a}. An uncharged static black ring solution is presented
in \cite{emparan:01}, but it contains conical singularities. Our
solution is the first asymptotically flat vacuum solution that is
completely regular on and outside an event horizon of non-spherical
topology. 

Our starting point is the following metric, constructed as
a Wick-rotated version of a solution in \cite{chamblin:97}:
\ba
\label{metric}
 ds^2 &=& -\frac{F(x)}{F(y)} \left(dt + \sqrt{\frac{\nu}{\xi_1}}
\frac{\xi_2-y}{A}
 d\psi \right)^2  \nonumber \\
      &+& \frac{1}{A^2 (x-y)^2} \left[ -F(x) \left( G(y) d\psi^2 +
 \frac{F(y)}{G(y)} dy^2 \right) 
\right. \\
 &+& \left. 
F(y)^2 \left(\frac{dx^2}{G(x)} +
\frac{G(x)}{F(x)} d\phi^2 \right) \right], \nonumber
\ea
where $\xi_2$ is defined below and
\be
 F(\xi) = 1 - \xi / \xi_1, \qquad G(\xi) = 1-\xi^2 + \nu \xi^3 \,.
\ee
The solution of \cite{chamblin:97} was obtained as the electric dual of the
magnetically charged Kaluza-Klein C-metric of \cite{dowker:94}. Our metric
can be related directly to the latter solution by analytic continuation. 
When $\nu=0$ we recover the static black ring solution of \cite{emparan:01}.


We assume that $0< \nu < \nu_* \equiv 2/(3 \sqrt{3})$, which ensures
that the roots of $G(\xi)$ are all distinct and real. They will be
ordered as $\xi_2<\xi_3< \xi_4$. It is easy to establish that $ -1 <
\xi_2 < 0 < 1 < \xi_3 < \xi_4 < \frac{1}{\nu}$. A double root
$\xi_3=\xi_4$ appears when $\nu = \nu_*$. Without loss of generality, we
take $A>0$. Taking $A<0$ simply reverses the sense of rotation.

We take $x$ to lie in the range $\xi_2 \leq x \leq \xi_3$ and
require that $\xi_1\geq \xi_3$, which ensures
that $g_{xx},g_{\phi\phi}\geq 0$. In order to avoid a conical
singularity at $x=\xi_2$ we identify $\phi$
with period
\be
\label{eqn:delphi}
\Delta \phi = \frac{4\pi \sqrt{F(\xi_2)}}{G'(\xi_2)}=
\frac{4\pi \sqrt{\xi_1- \xi_2}}{\nu \sqrt{\xi_1}(\xi_3 -
\xi_2) (\xi_4 - \xi_2)}.
\ee
A metric of Lorentzian signature is obtained by taking $y < \xi_2$.
Examining the behaviour of the constant $t$ slices of (\ref{metric}),
one finds that $\psi$ must be identified with period $\Delta \psi =
\Delta \phi$ in order to avoid a conical singularity at $y = \xi_2 \ne
x$. Regularity of the full metric here can be demonstrated by
converting from the polar coordinates $(y,\psi)$ to Cartesian
coordinates -- the $dt d\psi$ term can then be seen to vanish smoothly
at the origin $y = \xi_2$.

There are now two cases of interest depending on the value of
$\xi_1$. One of these will correspond to a black
ring and the other to the black hole of \cite{myers:86} with only one
non-vanishing angular momentum.

{\bf Case 1} is defined by $\xi_1 > \xi_3$.
In this case, $g_{\phi\phi}$ vanishes at $x=\xi_3$ and there will
be a conical singularity there unless $\phi$ is identified with period
\be
\label{eqn:xi3}
\Delta \phi' = \frac{4\pi \sqrt{F(\xi_3)}}{|G'(\xi_3)|}=
\frac{4\pi \sqrt{\xi_1-\xi_3}}{\nu \sqrt{\xi_1}(\xi_3
- \xi_2) (\xi_4 - \xi_3)}.
\ee
We demand $\Delta \phi = \Delta \phi'$ for consistency with
(\ref{eqn:delphi}). Since $\xi_2 < \xi_3$, this is possible only if
$\xi_1$ is fixed as a function of the other  three roots (i.e., of
$\nu$) as 
\be
\label{eqn:mudef}
 \xi_1 = \frac{\xi_4^2 - \xi_2 \xi_3}{2 \xi_4 - \xi_2 - \xi_3}
\qquad\mathit{(black \, ring)}.
\ee 
In this case it is easy to show that $\xi_3 < \xi_1 < \xi_4$,
from which it follows that the factors of $F(x)$ in the metric are
never zero. $x$ and $\phi$ parametrize a regular surface
of topology $S^2$.
The sections at constant $t,y$ have the topology of a ring $S^1\times
S^2$. $x=\xi_3$ is an axis pointing ``into'' the ring (i.e. decreasing
$S^1$ radius) and $x = \xi_2$ points out of the ring. This coordinate
system is sketched for the static black ring in \cite{emparan:01}; the
case considered here is very similar.

{\bf Case 2} is defined by
\be
\label{eqn:bhdef}
\xi_1=\xi_3 \qquad\mathit{(black \, hole)}\,.
\ee
In this case, $g_{\phi\phi}$ does not vanish at $x=\xi_3$, hence
(\ref{eqn:xi3}) need not be imposed. When (\ref{eqn:bhdef}) holds, the
sections at constant $t,y$ have the topology of three-spheres $S^3$,
with $\psi$ and $\phi$ being two independent rotation angles.

The following analysis applies to both case 1 and 2. 
Asymptotic infinity lies at $x =
y = \xi_2$. Defining $
 \tilde{\phi} = \frac{2\pi \phi}{\Delta \phi}$, $ \tilde{\psi} =
 \frac{2\pi \psi}{\Delta \psi}$,
and using the coordinate transformation
\be
 \zeta = \frac{\sqrt{\xi_2 - y}}{\tilde{A} (x-y)}, \qquad \eta =
 \frac{\sqrt{x-\xi_2}}{\tilde{A}(x-y)}\,,
\ee 
with ${\tilde{A}} = A\xi_1 \sqrt{\nu(\xi_3 -
 \xi_2) ( \xi_4 - \xi_2)}/ (2(\xi_1 - \xi_2))$,
the asymptotic metric is
brought to the manifestly flat form
\be
 ds^2 \sim -dt^2 + d\zeta^2 + \zeta^2 d\tilde{\psi}^2 + d\eta^2 +
\eta^2 d\tilde{\phi}^2 \,.
\ee
Note that the Killing vector fields ${\bf k} =
\partial/\partial t$, ${\bf m} = \partial/\partial
\tilde{\psi}$ are canonically normalized near infinity, and
$\tilde{\psi}$ and $\tilde{\phi}$ both have period $2\pi$.

The ADM mass and angular momentum are
\be
M = \frac{3\pi}{2GA^2} \frac{\xi_1- \xi_2}{\nu \xi_1^2(\xi_3 - \xi_2)
(\xi_4 -\xi_2)},
\ee
\be
J = \frac{2 \pi}{GA^3}\frac{(\xi_1 - \xi_2)^{5/2}}{\nu^{3/2}
\xi_1^3
(\xi_3 - \xi_2)^2(\xi_4-\xi_2)^2}.
\ee

The next limit to consider is $y \rightarrow -\infty$. 
Changing coordinates to $Y = -1/y$ gives a metric regular in a
neighbourhood of $Y=0$. Hence there is a new region $Y<0$, 
in which the coordinate $y$ can be defined as $y = -1/Y$ and the 
metric takes the same form as above.
The metric in this region is regular in these coordinates for $y>\xi_4$. 
${\bf k}$ becomes spacelike precisely at $Y=0$, so the
region $y > \xi_4$ will be referred to as the {\it ergoregion}. The
ergosurface at $Y=0$ has topology $S^1 \times S^2$ in case 1 and
$S^3$ in case 2. In both cases, ${\bf m}$ remains spacelike throughout 
the ergoregion.

The above coordinates break down at $y = \xi_4$, so define 
new coordinates $\chi$ and $v$ by
\be
 d\chi = d\psi + \frac{\sqrt{-F(y)}}{G(y)} dy,
\ee
\be
 dv = dt + \sqrt{\frac{\nu}{\xi_1}}(y - \xi_2)
 \frac{\sqrt{-F(y)}}{AG(y)} dy,
\ee
so $\chi$ is periodic with period $\Delta \chi = \Delta \psi$. 
In these new
coordinates, the metric takes the form
\ba
ds^2 &=& -\frac{F(x)}{F(y)} \left( dv - \sqrt{\frac{\nu}{\xi_1}}\frac{y
 - \xi_2}{A} d\chi \right)^2 \\
  &+& \frac{1}{A^2 (x-y)^2} \left[ F(x) \left( - G(y) d\chi^2 + 2
 \sqrt{-F(y)}d\chi dy \right) \right.\nonumber\\
 &+& \left. F(y)^2 \left(\frac{dx^2}{G(x)} +
\frac{G(x)}{F(x)} d\phi^2 \right) \right]. \nonumber
\ea
This is regular at $y = \xi_4$ so the
coordinate $y$ can now be continued into the region $y< \xi_4$. The
surface $y = \xi_4$ is a Killing horizon of the Killing vector field
\be
{\bf\xi} = \frac{\partial}{\partial v} + \frac{A\sqrt{\xi_1}}{\sqrt{\nu} 
 (\xi_4 -
 \xi_2)} \frac{\partial}{\partial \chi}
\ee
with surface gravity
\be
\kappa = \frac{A\sqrt{\nu}}{2}  \frac{\xi_1(\xi_4 -
\xi_3)}{\sqrt{
\xi_4 - \xi_1}}.
\ee 
Outside this horizon, $ {\bf \xi} = {\bf k} + \Omega_H  {\bf m}$,
where 
\be
 \Omega_H  = \frac{2 \pi A \sqrt{\xi_1} }{\Delta \phi(\xi_4 - \xi_2)
 \sqrt{\nu}}
= \frac{A\sqrt{\nu}}{2}  \frac{\xi_1(\xi_3 -
\xi_2)}{\sqrt{
\xi_1 - \xi_2}}.
\ee
Note that ${\bf \xi}$ is tangent to the null geodesic generators of
the horizon, and ${\bf \xi} \cdot \partial (\tilde{\psi}-\Omega_H t)
= 0$ on the horizon. It follows that the horizon is rotating with
angular velocity $\Omega_H $ with respect to the inertial frame at
infinity.

We have established that the solution possesses a rotating
horizon. The area of a constant time slice through the horizon is
\be
 {\cal A} = \frac{16 \pi^2}{A^3}\frac{(\xi_4 -
\xi_1)^{3/2}(\xi_1-\xi_2)}{\nu^{3/2}\xi_1^3 (\xi_4 - \xi_3)(\xi_3 -
\xi_2)(\xi_4-\xi_2)^2}.
\ee
In case 1, when the regularity condition (\ref{eqn:mudef})  is imposed,
the topology of (a constant time slice through) this event horizon is
$S^1 \times S^2$: it is a rotating black ring. In case 2, when 
(\ref{eqn:bhdef}) is imposed, the horizon is a rotating three-sphere.
The latter is actually the five-dimensional rotating black hole of
\cite{myers:86}, with one angular momentum parameter set to zero. To
see this, change coordinates to 
\be
r^2=\mu{(\xi_3-y)(\xi_4-x)\over (\xi_4-\xi_2)(x-y)}\,,\,
\cos^2\theta={(\xi_3-y)(x-\xi_2)\over
(\xi_3-\xi_2)(x-y)}\,,
\ee
and define
\be
\mu={4\over A^2\nu \xi_3^2(\xi_4-\xi_2)}\,,\quad
a={2\sqrt{\xi_3-\xi_2}\over A\sqrt{\nu}\xi_3(\xi_4-\xi_2)}\,,
\ee
(compare \cite{dowker:95}). Then one recovers the five-dimensional
Myers-Perry 
black hole in Boyer-Lindquist coordinates, $\mu$ and $a$ being the mass
and rotation parameters defined in \cite{myers:86}. We emphasize that
the above expressions for $M$, $J$, $\kappa$, $\Omega_H $ and ${\cal A}$
are valid for both the black ring and the black hole.

\begin{figure}[thb]
\begin{picture}(0,0)(0,0)
{\small 
\put(9,-7){0}
\put(106,-7){$\nu_*$}
\put(58,75){$\cal A$}
\put(55,30){$R_o$}
\put(58,10){$R_i$}
\put(124,-7){0}
\put(221,-7){$\nu_*$}
\put(133,50){$\kappa$}
\put(185,55){$\Omega_H$}
}
\end{picture}
\centering{\psfig{file=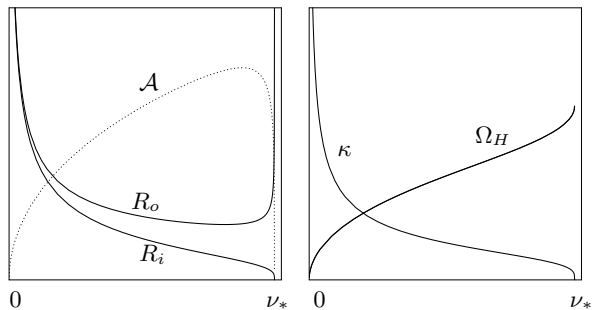,width=8cm}} 
\vspace{4ex}
\caption{Plots, as functions of $\nu$ at fixed $M$, of the inner
($R_i$) and outer ($R_o$) radii of curvature of the $S^1$, total area $\cal A$
of the ring, surface gravity $\kappa$, and angular velocity at the
horizon $\Omega_H$. All quantities are rendered dimensionless by
dividing by an appropriate power of $GM$.
}
\label{fig:doubleplot}
\end{figure}

Physically, (\ref{eqn:mudef}) is the condition that the rotation
balances the gravitational self-attraction of the ring, and it fixes a
relation between its mass, spin and radius. There are two independent
parameters for the solutions, $\nu$ and $A$.  $A$ has dimensions of
inverse length, and sets the scale for the solution. $\nu$ is
dimensionless and determines the shape of the $S^1 \times S^2$ horizon.
The $S^1$ can be  characterized by the inner radius of curvature $R_i$
at $x = \xi_3$,
and the outer radius $R_o$ at $x = \xi_2$. As $\nu \rightarrow 0$, both
radii tend to the same value $R\to 3J/M$ ($\to \sqrt{2}/A$). 
Also, $\Omega_H R\to 1$. Keeping $M$ fixed, the area
of the $S^2$ tends to zero  and $R$ tends to infinity, so small $\nu$
corresponds to a large thin ring.  
In fig.~\ref{fig:doubleplot} we plot physical quantities as a
function of $\nu$ when $M$ is fixed. As $\nu \to \nu_*$,
$\Omega_H  \to (3\pi/8GM)^{1/2}$. For $\nu$ near $\nu_*$ the ring is
highly flattened.

If $\nu=\nu_*$ then the black ring and the black hole
degenerate to the same solution with $\xi_3=\xi_1=\xi_4$. 
This is the $\mu = a^2$ limit of the five-dimensional rotating black
hole, for which the horizon disappears, and is replaced by a naked 
singularity. 

\begin{figure}[thb]
\begin{picture}(0,0)(0,0)
{\small 
\put(80,-8){$0.2164$}
\put(153,-8){$\nu_*$}
\put(-4,70){$1$}
\put(-25,58){$0.8437$}
\put(-5,-6){$0$}
}
\end{picture}
\centering{\psfig{file=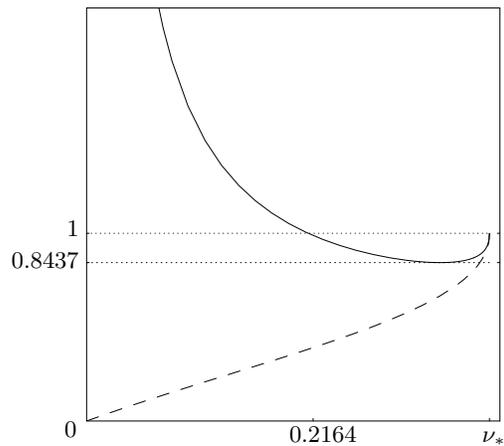,width=5.5cm}} 
\vspace{6ex}
\caption{$(27\pi/32G)J^2/M^3$ as a function of $\nu$. Here and in the
following graph, the solid line corresponds to the black ring, the
dashed line to the black hole. The two dotted lines delimit the values
for which a black hole and two black rings with the same mass and spin
can exist. 
}
\label{fig:JMplot}
\end{figure}

The spin of the five-dimensional black holes is bounded 
from above \cite{myers:86}:
\be
\label{eqn:Jbound}
 \frac{J^2}{M^3} \le \frac{32 G}{27 \pi} \,,
\ee
with equality for the (singular) $\mu = a^2$ solution. The
corresponding ratio for the black ring solutions is
\be
 \frac{J^2}{M^3} = \frac{32 G}{27 \pi} \frac{(\xi_4 -
 \xi_2)^3}{(2\xi_4 - \xi_2 - \xi_3)^2 (\xi_3 - \xi_2)}. 
\ee
These ratios are plotted as a function of $\nu$ in figure
\ref{fig:JMplot}. Note that the angular momentum of the ring is bounded
\textit{from below}, 
\be
\frac{J^2}{M^3} > 0.8437 \frac{32 G}{27 \pi}.
\ee
It is known that in six or more dimensions the spin of a black hole
can be arbitrarily large \cite{myers:86}. In five dimensions, we have
shown that the spin can also grow indefinitely, but only if the
spinning object is a ring!

For $0.2164 < \nu < \nu_*$, there are {\it two} black
ring solutions with the same values of $M$ and $J$ (but different
${\cal A}$). Moreover, these satisfy the bound (\ref{eqn:Jbound}) so
there is also a black hole with the same values of $M$ and $J$.
This is the first explicit demonstration that the uniqueness theorems 
valid in four dimensions do not have a simple generalization to five 
dimensions.

It is straightforward to check that the classical quantities
$M,J,\Omega_H,\kappa$ and ${\cal A}$ satisfy a Smarr relation
\be
 M = \frac{3}{2} \left(\frac{\kappa {\cal A}}{8\pi G} + \Omega_H J
 \right).
\ee
Another interesting formula relates $R_i$ to $M$ and $\kappa$:
\be
 \kappa = \frac{3\pi R_i}{8 G M}.
\ee
Since the temperature of the horizon is given by $\kappa/(2\pi)$, this
formula shows that the temperature of the ring is inversely
proportional to its mass per unit length (around the inner $S^1$).

If one considers perturbing the black ring in such a way that it
settles down to another black ring solution then the first law of
black hole mechanics can be proved in the usual way \cite{bardeen:73}
since this proof does not depend on the topology of the event
horizon. 

\begin{figure}[thb]
\begin{picture}(0,0)(0,0)
{\small
\put(30,-8){$0.918$}
\put(55,-8){$0.942$}
\put(92,-8){$1$}
\put(-5,-6){$0$}
\put(-20,60){$5.157$}
}
\end{picture}
\centering{\psfig{file=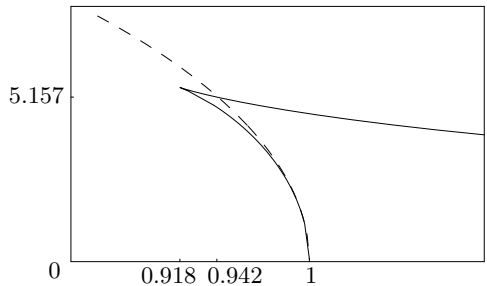,width=5.5cm}} 
\vspace{6ex}
\caption{${\cal A}/(GM)^{3/2}$ against $\sqrt{27\pi/32G}J/M^{3/2}$, around
the regime in which a black hole and two black rings with the same
$M$ and $J$ exist. For $\sqrt{27\pi/32G} J/M^{3/2}\approx 0.942$ there
exist a black hole and a black ring with the same mass, spin, and area
${\cal A}\approx 5.157 (GM)^{3/2}$.
}
\label{fig:areaplot}
\end{figure}

The solution with the larger area (entropy) for given values of $M$ and
$J$ is the one expected to be globally thermodynamically stable in the
microcanonical ensemble. In figure \ref{fig:areaplot} we have plotted
${\cal A}/M^{3/2}$ as a function of $J/M^{3/2}$. As the spin increases
(with the mass held fixed), there is first a small range of spins for
which the black hole has larger area than both black rings. However, at
a slightly larger spin, and before the singular limit is reached, the
larger black ring has a greater area than the black hole and will be the
preferred configuration. Hence we conjecture that, as a five-dimensional
black hole is spun up, a phase transition occurs from the black hole
to a black ring. The singular solution is never reached. 

Classically, the second law of black hole mechanics suggests that a
black hole might evolve into a black ring as angular momentum is added
to it. However, this involves a change in the topology of the horizon, 
and it is not clear whether this is possible classically (see
Ref.~\cite{horowitz:01} for an example in which classical topology
change of the horizon is forbidden). 

If $\nu \rightarrow 0$ at fixed $M$ then  the ring becomes large and
thin so one might expect ripples along the $\psi$ direction to lead to
a classical Gregory-Laflamme \cite{gregory:93} instability. However, if
the above conjecture is correct then we would expect a range of values
$0<\nu_1 < \nu <\nu_2 <\nu_*$ for which the ring is classically
stable.  The results of \cite{horowitz:01} suggest that the
horizon of unstable rings would tend to become lumpy around the $S^1$.
The changing quadrupole moment of such  an object would lead to
emission of  gravitational radiation until a stable endpoint was
reached, presumably either another ring or a spherical black hole
\cite{garyh}. 

The extremal limit of the black hole solution is a naked singularity,
and it appears that the black ring plays a role in smoothing out the
approach to this singularity. The existence of the ring may be
related to cosmic censorship. In four dimensions, the extremal limit of
the black hole is regular and the third law of black hole mechanics
forbids violation of the angular momentum bound $|J| \le GM^2$
\cite{wald:74,bardeen:73}. In dimension $D \ge 6$, black holes with a
single non-vanishing angular momentum can carry arbitrarily high $J$
(for a given $M$). Hence cosmic censorship would not require black rings
to exist in these cases. Five-dimensional black holes with {\it two}
non-zero angular momenta satisfy the bound (\ref{eqn:Jbound}) with $J$
replaced by $|J_1| + |J_2|$, but their extremal limit is non-singular so
the third law suggests that this bound cannot be violated by throwing
matter into the hole. However, one might wonder what would happen if one
were to throw some matter with $J_2 \ne 0$ into a black ring (with $J_2
= 0$). There may be a generalization of the black ring solution that
carries two angular momenta.

\medskip

RE acknowledges partial support from UPV grant 063.310-EB187/98 and
CICYT AEN99-0315. HSR was supported by PPARC. This paper is report
number CERN-TH/2001-294.

\end{document}